\pdfoutput=1 
\documentclass{IEEEcsmag}
\usepackage[frozencache=true,cachedir=minted-cache]{minted} 
\let\labelindent\relax
\usepackage{enumitem}
\usepackage{dirtree}
\usepackage{url}
\usepackage{lipsum}
\usepackage[colorlinks,urlcolor=blue,linkcolor=blue,citecolor=blue]{hyperref}
\expandafter\def\expandafter\UrlBreaks\expandafter{\UrlBreaks\do\/\do\*\do\-\do\~\do\'\do\"\do\-}
\usepackage{upmath,color}

\jvol{XX}
\jnum{XX}
\paper{8}
\jmonth{Month}
\jname{Computing in Science \& Engineering}
\jtitle{Publication Title}
\pubyear{2023}

\begin{document}

\sptitle{Department: Reproducible Research}
\title{Managing Software Provenance to Enhance Reproducibility in Computational Research}
   
\author{ Akash Dhruv, Anshu Dubey}
\affil{Argonne National Laboratory, Lemont, IL, 60439, USA} 

\markboth{DEPARTMENT}{DEPARTMENT}

\begin{abstract}
Scientific processes rely on software as an important tool for data acquisition, analysis, and discovery. Over the years sustainable software development practices have made progress in being considered as an integral component of research. However, management of computation-based scientific studies is often left to individual researchers who design their computational experiments based on personal preferences and the nature of the study. We believe that the quality, efficiency, and reproducibility of computation-based scientific research can be improved by explicitly creating an execution environment that allows researchers to provide a clear record of traceability. This is particularly relevant to complex computational studies in high-performance computing (HPC) environments. In this article, we review the documentation required to maintain a comprehensive record of  HPC computational experiments for reproducibility. We also provide an overview of tools and practices that we have developed to perform such studies around Flash-X, a multi-physics scientific software.

\end{abstract}

\maketitle

\section{INTRODUCTION}
Experimental and observational scientists follow a rigorous process of recording their work. For many, receiving training to do this diligently is considered a high priority for several excellent reasons. Unless the exact conditions of the experiment are recorded it cannot be repeated. A complete specification of assumptions made during the experiment, and the recording of the methodology followed for
interpreting and analyzing the outcome is required for
reproducibility -- a hallmark of the scientific
process. Additional factors may include the expense of running experiments, and the impact placement of probes might have on the recorded data.  In observational sciences, similar or greater constraints may result from the rarity of events -- the observer
must be ready to record the event when it occurs. To make the most of the event the observer would typically
wish to record as many details as possible.

This scientific discipline is largely lacking in computational science. It has traditionally been
assumed in computational science circles that careful recording of experiments may not be necessary because running the software again with identical input will produce identical results. This is a false assumption because software is rarely stationary. Every instance
of using it to obtain new scientific results tweaks it in some ways, thereby changing some of its behavior. In high-performance computing (HPC) 
 environments, the focus of this article,  changes in the system software stack and hardware may also cause
changes in results generated by the software. As the scale and complexity of the software and the problems it is trying to solve grow, so does the expense of running computational experiments. Some high-profile scientific failures in various disciplines and scrutiny of scientific output during COVID-19 have put a spotlight on the rigor of running and recording computational
experiments. Although workflow management tools have been developed to make it easier to run simulations, and archive and analyze the data, there is no community-wide adoption of rigor in requiring provenance of the code and data, and the log of the experiments conducted. 

Laboratory notebooks, or scientific notebooks (referred to as lab notebooks from here on) have long been used as a crucial tool in experimental and observational sciences by researchers, engineers, and students to record and document their work, experiments, observations, and findings in a structured and organized manner. Their purpose is to maintain a
detailed and accurate record of all activities related to a research project or experiment. Maintaining a well-kept lab notebook is not only considered a scientific best practice but also a requirement in
many research and academic settings. It helps ensure transparency, accountability, and the reliability of scientific work, which is
essential for advancing knowledge and innovation.

Individual researchers often use a directory-based organization for their computational experiments which are designed based on their personal preferences, the nature of the studies, and the need to balance time and computing resources. This directory-based design is helpful in nesting experiments that cover different parametric spaces. Leveraging this design and explicitly creating an execution environment around software and its dependencies can improve the quality, efficiency, and reproducibility of their studies. Such an environment, which is analogous to a laboratory space, can help focus researchers to prioritize scientific rigor and develop tools and practices to manage the generation, protection, storage, and analysis of data. The construction of this virtual laboratory environment should include the integration of a lab notebook that can provide a record of software configurations and research decisions.

 The question then is, what do lab notebooks look like for an HPC computational experiment, and how should they be managed? In this paper, we list features that lab notebooks may require for such computational investigations, and describe one exemplar of a  solution developed for conducting experiments with Flash-X \cite{DUBEY2022101168}, a multi-physics multi-component software that can be used for simulations in several science domains.

\section{LABORATORY NOTEBOOKS FOR COMPUTATIONAL EXPERIMENTS}
Since computational work often involves coding, simulations, and data
analysis using software tools, the lab notebook is best maintained in
electronic format, and could include a combination of repositories,
spreadsheets, text, and markdown formats. 
{For reasons discussed previously\footnote{\url{https://www.youtube.com/watch?v=fWpI4S_dvhc}}}, the use of some form of
lab notebook in computational science has several challenges.
{However, fundamental sets of activities that must be recorded are fairly
simple to enumerate}. 

\begin{itemize}
\item Title and Purpose: Title or project name and a brief description of its purpose or objectives.
\item Code Repository Links: Links or references to the code
repositories and/or input files used in the experiment.
\item  Software and Hardware: Specifications of hardware used and
system software stack, libraries, and tools used in the experiment,
along with their versions.
\item Modifications: Systematic recording of modifications to the software and
hardware during the experiment
\item Experiment Design: Description of the algorithms, data sets,
parameters, and any assumptions made for the experiment along with the
reasoning behind the selection of specific runs to be made. A log of steps
taken to prepare for the experiment can be very helpful for future experiments.
\item Data Sources: Documentation about any external data used, along with
explanations of how and where it's stored. Any data preprocessing
steps, such as cleaning, normalization, or transformation.
\item  Data Storage: Archival storage of produced and collected data with attached
metadata to be usable by other researchers.  
\item Experimental Runs: Log of each experimental run, including the
input parameters, the date and time of execution, and the resulting
output or data. If version control is in use commit references for any significant updates should
be kept.
\item Results and Analysis: Presentation of the results of
a computational experiment, including tables, graphs, and statistical
analyses along with a  description of how the results were interpreted and what conclusions were drawn.
\item References: Citations and references to relevant literature,
software documentation, or external resources that influenced the computational experiment.
\end{itemize}

Several tools and software applications can help maintain
well-organized and effective lab notebooks for computational
experiments. These tools are designed to streamline documentation,
code management, data analysis, and collaboration.  For example,
Jupyter Notebooks\footnote{\url{https://jupyter.org/}}  are interactive, web-based environments for creating
and sharing documents that combine live code (Python, R, etc.),
equations, visualizations, and narrative text. They are already widely used in
data science and computational research. Similarly R Markdown\footnote{\url{https://rmarkdown.rstudio.com/}} is an authoring
format that integrates code, results, and narrative text into a single
document. It's commonly used with the R programming language but can
be adapted for other languages as well. 
Python-based notebooks such as Google Colab\footnote{\url{https://colab.google/}} are similar to Jupyter Notebooks but are focused on Python. They allow documentation of Python code alongside explanations and visualizations.
Several other open-source and commercial solutions exist with a variety of features. Any of these tools along with a set of well-defined recording practices can form the basis of an execution environment that promotes reproducibility of computational experiments. However, none of them suffice for complex multiphysics HPC computations which generate a huge amount of data, and require substantial post-processing and analysis.

\section {THE FLASH-X SOLUTION} \label{sc:tooling}
Flash-X is community-developed software that is undergoing several modes of development simultaneously. It is a new version of a long-existing community code FLASH \cite{DUBEY2009} that has been re-architected to be compatible with heterogeneous hardware platforms. Several new physics capabilities and an entirely new method of integration have also been added to the code.  As a consequence, a fairly common occurrence is where ongoing capability and performance improvement requirements collide with the needs of a domain science study. The situation is exacerbated when all participants involved in a study are not familiar with the inner workings and constraints of the code. In general two different types of experiments are regularly conducted with the code. One set of experiments measures the performance of different components of the code, while the other set pertains to domain science investigations. The requirements and constraints of these experiments differ from one another, though some of the required meta-information for complete specification is identical for all experiments. Some of this required meta-information such as repository version, software stack version, and all the configuration parameters are recorded in a log file that every execution instance of Flash-X generates.

The performance experiments typically involve scaling studies with the implication that the same application is run with different configurations of degree of parallelism, hardware components in use, and possibly different implementations of some of the code components. The exploration parameters in such studies tend to be related to the infrastructural components of the code, and the analysis is performed on either the performance summary section of the log file or, if a performance tool is used, then the data generated by the tool. The output of the simulation itself is not relevant to the study. The domain science investigations explore the parameter space of the physics involved. These experiments typically have little variation in the degree of parallelism or hardware in use. The output of the simulation is an important artifact here and can be quite large. 

\begin{figure}[h]
\centerline{\includegraphics[width=18.5pc]{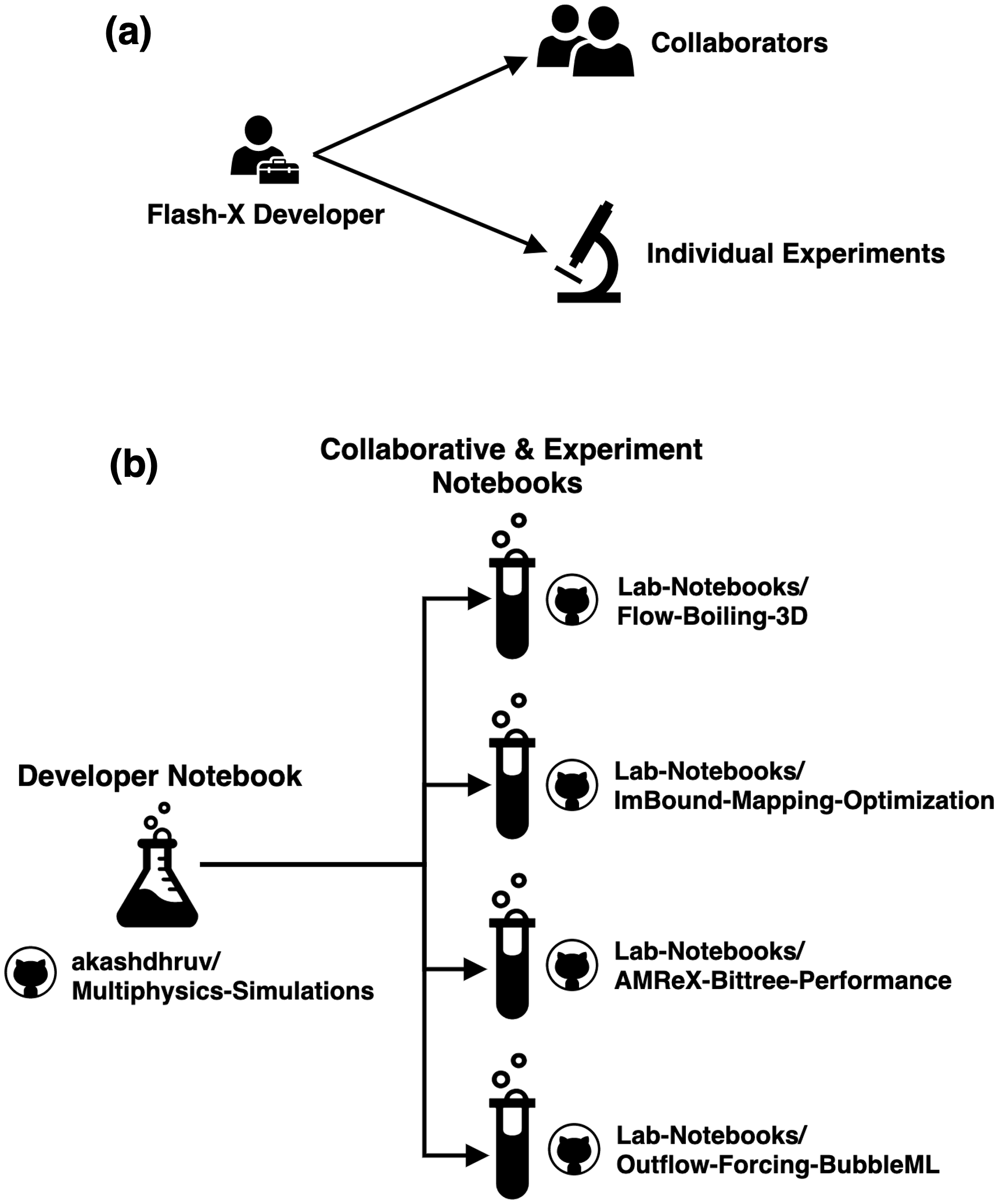}}
\caption{(a) A common scenario in computational sciences where a developer has to conduct individual experiments and collaborate with other researchers who may not be experts in using their software. (b) Example use case of lab notebooks for Flash-X development. The developer maintains a personal repository that serves as a seed for setting up experiments for physics and performance related studies that can be shared with collaborators.}
\label{fig:case-study-example}
\end{figure}

An example of a workflow that might be encountered by a developer who is also a user of the code for doing scientific investigations is shown in Figure \ref{fig:case-study-example}a. Such developers may perform regular testing of the physics and infrastructure components, keep track of the performance changes, and conduct scientific experiments at the same time.  {The collection of lab notebooks \cite{Lab-Notebooks} described in Figure \ref{fig:case-study-example}b provides a good example of what may be done in such circumstances}. These notebooks are all seeded from a general notebook \cite{Multiphysics-Simulations} being maintained by the developer in question. In this instance, the developer is simultaneously conducting the following experiments,

\begin{itemize}
    \item \textbf{Flow-Boiling-3D}: Lab notebook for production runs of three-dimensional multiphase flow boiling simulations.
    \item \textbf{ImBound-Mapping-Optimization}: Lab notebook to investigate and optimize mapping of Lagrangian particles on block-structured AMR grids for the purpose of developing an immersed boundary method for fluid-structure interaction problems.
    \item \textbf{AMReX-Bittree-Performance}: Lab notebook to improve scaling of Flash-X applications that use AMReX in octree mode.
    \item \textbf{Outflow-Forcing-BubbleML}: Reproducibility capsule for research articles on using Flash-X simulations to develop scientific machine learning models for thermal science applications. 
\end{itemize}

\begin{figure}
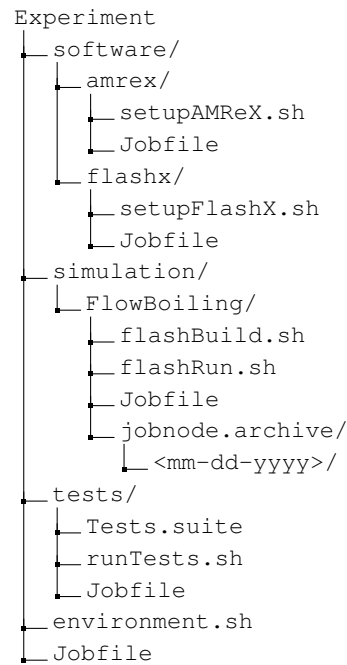

\dirtree{%
.1 Experiment.
.2 software/.
.3 amrex/.
.4 setupAMReX.sh.
.4 Jobfile.
.3 flashx/.
.4 setupFlashX.sh.
.4 Jobfile.
.2 simulation/.
.3 FlowBoiling/.
.4 flashBuild.sh.
.4 flashRun.sh.
.4 Jobfile.
.4 jobnode.archive/.
.5 <mm-dd-yyyy>/.
.2 tests/.
.3 Tests.suite.
.3 runTests.sh.
.3 Jobfile.
.2 environment.sh.
.2 Jobfile.
}
\caption{Directory tree for a Flash-X experiment.}
\label{fig:dirtree}
\end{figure}

Figure \ref{fig:dirtree} depicts the directory tree that might be set up to record this set of experiments. Here, the first subdirectory, {\em software} records all the software configuration information that is not directly available from the log file. In this example there is an external dependency AMReX \cite{AMReX_JOSS}, a library that provides adaptive mesh refinement capabilities. The build specifications of external libraries are not known to Flash-X configurator, and must be explicitly recorded to be able to exactly reproduce the execution environment. The second subdirectory {\em simulation} describes the application instance that is used in the experiment. If an experiment uses more than one application instance it is expected that a separate subdirectory would be created for each. Because Flash-X has a high degree of composability, and therefore a huge collection of tests, it is also important to list the tests that were used to ensure ongoing correctness of the code throughout the duration of the experiment. Organizing the tests within this structure also enables quick running of the test-suite as described later. 

The process of seeding and managing these notebooks is implemented using a lightweight command line tool, Jobrunner \cite{Jobrunner}, that enforces a directory-based inheritance for application configurations encoded in shell scripts to manage computational experiments. The directory tree is designed to split different shell commands that set up the software stack and simulation in an optimal way. For instance, a combination of {environment.sh} and {setupAMReX.sh} is used to configure AMReX most suitably for the experiment. Here, {environment.sh} documents and sets variables that store compiler information and installation location of different libraries, and {setupAMReX.sh} stores information specific to building AMReX with the necessary options. { The {Jobfiles} that exist at every level of the tree encode information that Jobrunner can use to stitch together these bash files to perform different tasks.} 
\begin{figure}
\begin{minted}[
    gobble=1,
    frame=single,
  ]{yaml}
  # Location:
  # Experiment
  job:
    setup:
        - environment.sh
    submit:
        - environment.sh

  # Location:
  # Experiment/software/amrex
  job:
    setup:
        - setupAMReX.sh

  # Location: 
  # Experiment/simulation/FlowBoiling
  job:
    setup:
        - flashSetup.sh
    submit:
        - flashSubmit.sh
    archive:
        - "*hdf5*"
        - "*.log"
\end{minted}
\caption{Contents of {Jobfiles} at different locations along the directory tree.}
\label{fig:jobfiles}
\end{figure}

Figure \ref{fig:jobfiles} shows contents of the {Jobfiles} at different nodes along the directory tree described in Figure \ref{fig:dirtree}. During the execution of Jobrunner commands, files assigned to the respective commands are picked up to execute tasks in the target node of the directory tree. For instance, the commands described in Figure \ref{fig:commands} build AMReX, clone and configure an application instance of Flash-X for multiphase pool boiling problem, compile the assembled code, and then execute the created binary to obtain results. At the same time, local testing of the code is performed using Flash-X's custom testing framework\footnote{\url{https://github.com/Flash-X/Flash-X-Test}} which uses tests encoded in {Tests.suite} along with the {environment.sh} and {runTests.sh} scripts to build and execute tests. Note that environment.sh, located at the root of the project directory is used for each Jobrunner command, providing consistency between tests and experiments. 

Jobrunner hides error-prone individual steps of the experiments while allowing organized and explicit documentation of configuration options that can be easily modified by editing the shell scripts. The directory tree in Figure \ref{fig:dirtree} can be easily modified and redesigned based on the requirements of the experiments to include markdown notes and analysis files. {For full functionality and documentation for Jobrunner see A. Dhruv \cite{Jobrunner}.}
\begin{figure}
\begin{minted}[
    gobble=8,
    frame=single,
  ]{bash}
        # Setting up software and dependencies
        jobrunner setup software/amrex
        jobrunner setup software/flashx

        # Setting up and running experiments
        jobrunner setup simulation/FlowBoiling
        jobrunner submit simulation/FlowBoiling

        # Archiving results to jobnode.archive
        jobrunner archive \
                    simulation/FlowBoiling

        # Running Flash-X test suite
        jobrunner submit tests
\end{minted}
\caption{Jobrunner commands for setting up dependencies, running tests and experiments, and archive data. These commands are executed from the root of the directory-tree}
\label{fig:commands}
\end{figure}

Data archiving is the last remaining concern for reproducibility. This is implemented using Jobrunner's archive command which picks up file patterns listed in Jobfiles (see Figure \ref{fig:jobfiles}) and moves them to a jobnode.archive/<mm-dd-yyyy> directory under the target node of an experiment (see Experiment/simulation/FlowBoiling in Figure \ref{fig:dirtree}). The directories containing the data are eventually moved to a cloud-based archival service along with a clone of the Github source repository to preserve the tree structure. Note that the raw data is itself not included in the repository because that can be quite large. It is organized in such a way that the archive can be unpacked below the directory structure maintained in the repository to exactly reconstruct all the artifacts of the experiment.  
See Lab-Notebooks/Outflow-Forcing-BubbleML  \cite{Lab-Notebooks} for an example lab notebook for recent publications. 

\section {CONCLUSION} \label{sc:conclusion}
An increasing emphasis on reproducibility and greater scrutiny of computational science results is slowly changing the perception of what constitutes a good computational experiment. The concept of maintaining laboratory notebooks in computational sciences has been gaining popularity to enforce structure and rigor in scientific studies. This is a welcome change, and deserves encouragement. However, unlike more traditional scientific disciplines, computational scientists have challenges in how frequently their execution environments may change, and the kind of impact they may have on the continuity of their experiments. Additionally, they often participate in teams that may be geographically diverse, therefore they need distributed digital mechanisms to record their work. Tools such as Jupyter notebook and Github repositories can help, but need additional care to fully capture the provenance of a computational experiment. We have presented an approach that addresses many of the challenges faced by software that is being developed while also being used for production. It is our hope that tools and ideas presented here serve as a motivation for other scientists to design and organize their experiments. Integration of well-organized laboratory notebooks with reproducibility and data capsules can improve the quality of scientific artifacts, and enhance productivity of collaborative research.

\section {ACKNOWLEDGMENTS}

The authors would like to acknowledge Jared O'Neal's work in promoting the culture of using lab notebooks for Flash-X experiments, and for conceptualizing the format that has been adopted by the team and described in this article. 

This work was partially supported by the Exascale Computing Project (17-SC-20-SC), a collaborative effort of the US Department of Energy Office of Science and the National Nuclear Security Administration, and the Laboratory Directed Research and Development Program supported by Argonne. The submitted manuscript was created by UChicago Argonne, LLC, operator of Argonne National Laboratory (“Argonne”). Argonne, a U.S. Department of Energy Office of Science laboratory, is operated under Contract No. DE-AC02-06CH11357. The U.S. Government retains for itself, and others acting on its behalf, a paid-up nonexclusive, irrevocable worldwide license in said article to reproduce, prepare derivative works, distribute copies to the public, and perform publicly and display publicly, by or on behalf of the Government. The Department of Energy will provide public access to these results of federally sponsored research in accordance with the DOE Public Access Plan. http://energy.gov/downloads/doe-public-access-plan.

\begin{IEEEbiography}{Akash Dhruv,} is a Postdoctoral Fellow in the Mathematics and Computer Science Division at Argonne National Laboratory where he performs research in DevOps and performance engineering for scientific computing applications, along with building computational pipelines to integrate simulation with machine learning workflows. He received his Ph.D. in mechanical and aerospace engineering from George Washington University, Washington D.C., and a B.Tech in mechanical engineering from National Institute of Technology, Surat. \vspace*{8pt}
\end{IEEEbiography}

\begin{IEEEbiography}{Anshu Dubey,} is a Senior Computational Scientist in the Mathematics and Computer Science Division at Argonne National Laboratory and a Senior Scientist
(CASE) at the University of Chicago. She was previously at Lawrence Berkeley National Laboratory, and before that she was the associate director of the Flash Center for Computational Science at the University of Chicago and also led the CS/Applications
group, who develop, maintain, and distribute the FLASH code. She
received her Ph.D. in computer science from Old Dominion University and
a B.Tech. in electrical engineering from the Indian Institute of Technology,
New Delhi.\vspace*{8pt}
\end{IEEEbiography}

\end{document}